\documentclass[11pt]{article}
\begin{document}

\begin{center}
\large\textbf{Modifying horizon thermodynamics by surface tensions}
\end{center}

\begin{center}
Deyou Chen$^{a}$\footnote{ E-mail: \underline
{dchen@cwnu.edu.cn} } and Xiaoxiong Zeng$^{b,c}$\footnote{ E-mail:
\underline {xxzeng@itp.ac.cn} }
\end{center}

\begin{center}
$^{a}$College of Physics and Space Science, China West Normal University, \\Nanchong 637009, China

$^{b}$School of Material Science and Engineering, Chongqing Jiaotong University, Chongqing, 400074, China

$^{c}$Key Laboratory of Frontiers in Theoretical Physics, Institute of Theoretical Physics, Chinese Academy of Sciences, Beijing 100190, China
\end{center}

\textbf{Abstract:}
The modified first laws of thermodynamics at the black hole horizon and the cosmological horizon of the Schwarzschild de Sitter black hole and the apparent horizon of the Friedmann-Robertson-Walker cosmology are derived by the surface tensions, respectively. The corresponding Smarr relations are obeyed. For the black hole, the cosmological constant is first treated as a fixed constant, and then as a variable associated to the pressure. The law at the apparent horizon takes the same form as that at the cosmological horizon, but is different from that at the black hole horizon. The positive temperatures guarantee the appearance of the worked terms in the modified laws at the cosmological and apparent horizons. While they can disappear at the black hole horizon.

\section{Introduction}
Einstein's equations can be written as a thermodynamic identity. It was put forward by Padmanabhan and first proved in the spherically symmetric spacetime where Einstein's equations near the horizon are in the form of the first law of thermodynamics

\begin{equation}
\delta E=T\delta S-P\delta V. \label{eq1.1}
\end{equation}

\noindent In the above equation, $E$ is the energy of the spacetime geometry, $P$ is the pressure provided by the source of the Einstein's equations and $\delta V$ is the change of the volume \cite{TP,RGC}. Subsequently, Sarkar and Kothawala found that this view is also applicable to the spherically symmetric horizons of various theories of gravity and the tunneling rate $\Gamma \sim e^{\Delta S}$ is a natural consequence of the first law \cite{SSDK}. Near the evolving spherically symmetric horizons and the stationary axisymmetric horizons, the laws were obtained from the Einstein's equations in the references \cite{TK,KSP,AS}. Other researches on the connection between horizon thermodynamics and gravitational dynamics are referred to \cite{CP,CPP,SC} and the references therein. All of these researches effectively support the view of Padmanabhan. In the anti de Sitter spacetime, Kastor et al first treated the cosmological constant as the pressure $P=-\frac{\Lambda}{8\pi}$ and its conjugate quantity as the thermodynamic volume, thereby the mass of an AdS black hole is interpreted as the enthalpy of the spacetime \cite{KRT}. Adopting this view, Kubiznak and Mann studied the thermodynamics of the charged AdS black hole in the extended phase space \cite{KM}. The first law was gotten as

\begin{equation}
\delta M=T\delta S+\Phi \delta Q  +V\delta P, \label{eq1.2}
\end{equation}

\noindent which obeys the corresponding Smarr relation, where $M$ is the black hole mass identified as the enthalpy.  They found that the critical exponents of the black hole are full in consistence with those of the Van der Waals system. This work is the further elaboration of Dolan's work where the critical behaviour of the AdS black holes in the extended phase space was discussed and the analogy with the Van der Waals was found \cite{BPD}. Based this interesting work, the thermodynamics of various complicate spacetimes were discussed and many significative critical phenomena were found \cite{GKM,WL,WL1,HV,CCLY,AKM,ZZW,GKY,CVJ,BMS,STK,DKS,XZ,MMR,AOS,ZJ,ML,PMS}.

Recently, Hansen et al found that the radial Einstein equation at the horizon of the Kerr black hole results in the modified first law of thermodynamics

\begin{equation}
\delta E=T\delta S+\Omega \delta J  - \sigma \delta A, \label{eq1.3}
\end{equation}

\noindent where $E$ is the Misner-Sharp mass, $J=Ea$ is the horizon angular momentum, $\sigma$ denotes the surface tension at the horizon and $A$ is the horizon area \cite{HKM}. In this equation, the horizon radius $r_+$ and the rotation parameter $a$ are two independent variables, which is different from Eq. (\ref{eq1.1}). Using the relation between the area and the volume and ordering $J=0$, one can reduce it to Eq. (\ref{eq1.1}). However, the reduced equation doesn't obey the corresponding Small relation.

Our aim in this paper is to investigate the thermodynamics at the black hole horizon and the cosmological horizon of the Schwarzschild de Sitter black hole and at the apparent horizon of the Friedmann-Robertson-Walker (FRW) cosmology by the surface tensions. The modified first laws of thermodynamics are gotten and obey the corresponding Smarr relations. For a de Sitter black hole contains two horizons, the temperatures are different at different horizons. Therefore, the nonequilibrium state exists and it is difficult to discuss the thermodynamics. To overcome this difficulty, several approaches were adopted. One approach is defining a mass-energy-like quantity at infinity in asymptotically spacetimes \cite{GM,BBM}. The second way is treating the black hole horizon and the cosmological horizon as two independent thermodynamical systems \cite{UTS1,YS,AC,CJS,WH}. Besides, the global approach is the construction of the globally effective temperature and other effective thermodynamic quantities \cite{UTS,ZZMZ}. Here we adopt the second way to investigate the thermodynamics. We first let the cosmological constant be fixed and derive the modified first laws at the black hole horizon and the cosmological horizon. Then treating the constant as a variable associated to the pressure $P=-\frac{\Lambda}{8\pi}$, we get the laws in the extended phase spaces. Considering the pure de Sitter spacetime is a special case of the FRW spacetime, we formulate the extension to the FRW cosmology and obtain the modified first law at the apparent horizon. The modified law at the apparent horizon has the same form as that at the cosmological horizon. However, it is different from that at the black hole horizon.

The rest is organized as follows. In the next section, we derive the modified first laws of thermodynamics at the black hole horizon by the surface tension. The cosmological constant is treated as a fixed constant and a variable, respectively. In section 3, the modified laws at the cosmological horizon are gotten. In section 4, the modified law at the apparent horizon of the FRW cosmology are investigated. Section 5 is devoted to our discussion and conclusion.

\section{Modified thermodynamics at the black hole horizon}
The thermodynamics of the Schwarzschild de Sitter black hole were studied in \cite{GM,BBM,YS,UTS,GT}. The conserved quantities defined by integrals were introduced to obtain the first law of thermodynamics. However, the problem of integration constant exists. To overcome this problem, the Iyer-Wald formalism was adopted \cite{UTS}. The cosmological constant was seen as a fixed constant in the former discussions \cite{GM,BBM,GT}. Subsequently, it was treated as an independent variable in \cite{YS,UTS}.

The Schwarzschild de Sitter black hole is given by

\begin{equation}
ds^2 = -f(r) dt^2 + \frac{1}{f(r) } dr^2 +r^2(d\theta^2 +sin^2{\theta}d{\phi}^2), \label{eq2.1}
\end{equation}

\noindent where $f(r) =1- \frac{2M}{r}-\frac{\Lambda r^2}{3}$, $\Lambda$ is the cosmological constant and $M$ is the physical mass. There are two positive roots and a negative root for $f(r) =0$. Our discussion is limited to the range $0<9\Lambda M<1$ which guarantees the existence of the black hole horizon $r_+$ and the cosmological horizon $r_C$. We first investigate the thermodynamics at the black hole horizon by the surface tension. The entropy and the temperature at the black hole horizon are

\begin{eqnarray}
S_+ &=& \pi r_+^2, \quad T_+ = \frac{f^{\prime}(r_+)}{4\pi}= \frac{1-\Lambda r_+^2}{4\pi r_+}, \label{eq2.2}
\end{eqnarray}

\noindent respectively, where  $f^{\prime}(r_+)=\frac{\partial f(r)}{\partial r}\mid _{r=r_+} $. There are several definitions of mass. In this paper, we adopt the definition of the Misner-Sharp mass. Using the definition \cite{MS,CM,SAH,HZ}, we get the Misner-Sharp mass surrounded by the black hole horizon as

\begin{equation}
E_+= \frac{r_+}{2}. \label{eq2.3}
\end{equation}

To derive the surface tension and the thermodynamical identity, we first calculate the radial Einstein equation at the black hole horizon and get

\begin{equation}
G_r^r\mid_{r_+} = 8\pi T_r^r\mid_{r_+}= \frac{r_+ f^{\prime}(r_+)-1}{r_+^2}. \label{eq2.4}
\end{equation}

\noindent Solving the above equation yields

\begin{equation}
f^{\prime}(r_+) = 8\pi r_+ T_r^r\mid_{r_+}+\frac{1}{r_+}. \label{eq2.5}
\end{equation}

\noindent Moving the first term on the right hand side (rhs) of the above equation to the left hand side (lhs) and multiplying by $\frac{\delta S_+}{4\pi}$ on the both sides of the moved equation yield

\begin{equation}
\frac{f^{\prime}(r_+)}{4\pi}\delta S_+ - 2 r_+ T_r^r\mid_{r_+}\delta S_+ = \frac{\delta r_+}{2}. \label{eq2.6}
\end{equation}

\noindent Here we have used the differential expression $\delta S_+ = 2\pi r_+ \delta r_+$ to derive the term on rhs of Eq. (\ref{eq2.6}). It is clearly that $\frac{f^{\prime}(r_+)}{4\pi}$ is the value of the temperature at the black hole horizon. So the first term on lhs is in of the form $T_+ \delta S_+$. The term $\frac{1}{2}\delta r_+$ equals $\delta E$ and is identified as the change of the energy. $2 r_+ T_r^r\mid_{r_+}\delta S_+$ is written as $\sigma_+\delta A_+$ by using the relation between the entropy and the horizon area $A_+$. Therefore, Eq. (\ref{eq2.6}) is written as

\begin{equation}
\delta E_+= T_+ \delta S_+  -\sigma_+ \delta A_+, \label{eq2.7}
\end{equation}

\noindent which is the modified first law of thermodynamics at the black hole horizon and $\sigma_+ = \frac{1}{2} r_+ T_r^r\mid_{r_+}=-\frac{\Lambda r_+}{16\pi}$ denotes the surface tension. Clearly, the corresponding Smarr relation

\begin{equation}
E_+= 2(T_+ S_+  -\sigma_+ A_+) \label{eq2.8}
\end{equation}

\noindent is obeyed by using the expressions of $T_+,S_+,\sigma_+$ and $A_+$. The Gibbs free energy at the horizon in the de Sitter spacetime is

\begin{equation}
G_+ = E_+ - T_+ S_+ + \sigma_+ A_+, \label{eq2.9}
\end{equation}

\noindent which obeys differential expression $\delta G_+ = -S_+\delta T_+ + A_+\delta \sigma_+$. After a simple calculation, the energy is gotten as $G_+=\frac{r_+}{4}$. When Eq. (\ref{eq2.7}) is reduced to $\delta E_+= T_{+e} \delta S_+$ by the relation between the entropy and the area, $T_{+e}=T_+-4\sigma_+ = \frac{1}{4\pi r_+}$ expresses the effective temperature at the black hole horizon.

Taking into account the cosmological constant as a variable associated to the pressure in the recent work, we make further to explore the thermodynamics and write Eq. (\ref{eq2.7}) as

\begin{equation}
\delta E_+= T_+ \delta S_+  -P_+ \delta V_+, \label{eq2.10}
\end{equation}

\noindent by using the relation between the area $\sigma_+$ and the volume $V_+$, where $P_+ =\frac{2\sigma_+}{r_+}=-\frac{\Lambda}{8\pi}$ expresses the pressure at the horizon.  Eq. (\ref{eq2.10}) can be furthermore written as

\begin{equation}
\delta E_{0+}= T_+ \delta S_+  +V_+ \delta P_+ , \label{eq2.11}
\end{equation}

\noindent where $ E_{0+} = E_{+}+P_+V_+=M$. The corresponding Smarr relation $E_{0+}= 2(T_+ S_+  -P_+V_+)$ is also satisfied when the related expressions are introduced. Therefore, Eq. (\ref{eq2.11}) is the first law of thermodynamics at the black hole horizon in the extended phase space. So both of Eq. (\ref{eq2.7}) and Eq. (\ref{eq2.11})) are the first law at the black hole horizon.

\section{Modified thermodynamics at the cosmological horizon}

In this section, we investigate the thermodynamics at the cosmological horizon of the Schwarzschild de Sitter black hole by the surface tension. A parallel process as the section 2 is performed. The temperature at this horizon is lower than that at the black hole horizon. The  energy of a positive mass at the cosmological horizon is measured as a negative value. The entropy and the temperature at the cosmological horizon are

\begin{eqnarray}
S_C &=& \pi r_C^2, \quad T_C = \frac{\mid f^{\prime}(r_C) \mid}{4\pi}= -\frac{ f^{\prime}(r_C)}{4\pi}, \label{eq3.1}
\end{eqnarray}

\noindent respectively, where  $f^{\prime}(r_C)=\frac{\partial f(r)}{\partial r}\mid _{r=r_C} = \frac{1-\Lambda r_C^2}{ r_C}<0$. The Misner-Sharp mass surrounded by the cosmological horizon is gotten as

\begin{equation}
M_C= \frac{r_C}{2}. \label{eq3.2}
\end{equation}

To investigate the surface tension and the modified horizon thermodynamics, we first calculate the radial Einstein equation at the cosmological horizon and get

\begin{equation}
G_r^r\mid_{r_C} = 8\pi T_r^r\mid_{r_C}= \frac{r_C f^{\prime}(r_C)-1}{r_C^2}, \label{eq3.3}
\end{equation}

\noindent which yields

\begin{equation}
f^{\prime}(r_C) = 8\pi r_C T_r^r\mid_{r_C}+\frac{1}{r_C}. \label{eq3.4}
\end{equation}

\noindent Multiplying by $-\frac{\delta S_C}{4\pi}$ on the both sides of the above equation and moving the first term of rhs to lhs, we obtain

\begin{equation}
-\frac{f^{\prime}(r_C)}{4\pi}\delta S_C + 2 r_C T_r^r\mid_{r_C}\delta S_C = -\frac{\delta r_C}{2}, \label{eq3.5}
\end{equation}

\noindent where $-\frac{f^{\prime}(r_C)}{4\pi}$ is the temperature at the cosmological horizon and the first term on lhs can be written as $T_C dS_C$. The differential expression $\delta S_C = 2\pi r_C \delta r_C$ is introduced to derive the term on rhs of Eq. (\ref{eq3.5}). $ \delta M_C =\frac{1}{2}\delta r_C$ expresses the change of the Misner-Sharp mass at the cosmological horizon. We identify the negative Misner-Sharp mass as the energy $E_C$ and then obtain $\delta E_C = -\frac{\delta r_C}{2}$ which shows the change of energy is related to that of the horizon location \cite{TP}. The second term $2 r_C T_r^r\mid_{r_C}\delta S_C$ is written in the form of $\sigma_C \delta A_C$ by the relation between the entropy and the horizon area, where $\sigma_C = \frac{1}{2} r_C T_r^r\mid_{r_C}=-\frac{\Lambda r_C}{16\pi}$. The inner of a black hole horizon is defined as an inaccessible region surrounded by the horizon. The horizon area increases when a particle is absorbed by the black hole, and then the change of the horizon area is positive ($\delta A_+ >0$). While the outer of the cosmological horizon is an inaccessible region. When a particle enter the cosmology, the volume increases with the increase of the horizon area and radius. Thus, Eq. (\ref{eq3.5}) is rewritten as

\begin{equation}
\delta E_C= T_C \delta S_C  +\sigma _C \delta A_C, \label{eq3.6}
\end{equation}

\noindent which is the modified first law of thermodynamics at the cosmological horizon and $\sigma$ is the surface tension. The corresponding Smarr relation

\begin{equation}
E_C= 2(T_C S_C  +\sigma _C A_C) \label{eq3.7}
\end{equation}

\noindent is obeyed. The Gibbs free energy at this horizon is

\begin{equation}
G_C = E_C- T_C S_C - \sigma _C A_C. \label{eq3.8}
\end{equation}

\noindent Using the related expressions, we get $G_C=-\frac{r_C}{4}$. The differential expression $\delta G_C = -S_C\delta T_C - A_C\delta \sigma_C$ is also satisfied. From the relation between the entropy and the area, Eq. (\ref{eq3.6}) can be written as $\delta E_C= T_{Ce} \delta S_C$ which shows that the change of the energy is completely caused by absorbing (or releasing) heat. $T_{Ce}=T_C +4\sigma_C = -\frac{1}{4\pi r_C}$ denotes the effective temperature. However, this process is not allowed due to the appearance of the negative temperature. When the energy changes, both of the following phenomena happen simultaneously, namely, the system is doing work to the external world and the heat is absorbed or released. Therefore, the worked term $\sigma _C \delta A_C$ must exist in the first law at the cosmological horizon.

Now we treat the cosmological constant as the pressure and reconsider the thermodynamics. Using the relation between the area $\sigma_C$ and the volume $V_C$, we write Eq. (\ref{eq3.6}) as

\begin{equation}
\delta E_C= T_C \delta S_C  +P_C \delta V_C, \label{eq3.9}
\end{equation}

\noindent where $P_C =-\frac{\Lambda}{8\pi}$ denotes the pressure at the cosmological horizon. Eq. (\ref{eq3.9}) can be furthermore written as

\begin{equation}
\delta E_{0C}= T_C \delta S_C  -V_C \delta P_C , \label{eq3.10}
\end{equation}

\noindent where $ E_{0C} = E_{C}-P_CV_C=-M$. This formula is the first law of thermodynamics in the extended phase space. The corresponding Smarr relation $E_{0C}= 2(T_C S_C  + P_CV_C)$ is also satisfied. Therefore, both of the formulae (\ref{eq3.6}) and (\ref{eq3.10}) are the modified first laws of thermodynamics at the cosmological horizon.

\section{Modified thermodynamics at the apparent horizon of the FRW cosmology}

The modified thermodynamics at the apparent horizon of the FRW cosmology is derived by the surface tension in this section. The thermodynamics were studied in \cite{CK,GW,CCH1,LZ,ZRS}. In \cite{CK}, Cai and Kim derived the Friedmann equations describing the dynamics of the cosmology with any spatial curvature from the first law of thermodynamics at the apparent horizon and the Bekenstein area-entropy formula. The law is expressed as $-dE=T dS$, where $dE$ is related to the Misner-Sharp mass and expresses the energy crossing the apparent horizon during an infinitesimal time interval. On the other hand, from the Friedmann equation, the first law at the apparent horizon was gotten in \cite{GW}.

The FRW metric is given by

\begin{equation}
ds^2 = - dt^2 + a^2(t)\left(\frac{dr_0^2}{1-kr_0^2} +r_0^2d\Omega_2^2\right), \label{eq4.1}
\end{equation}

\noindent where $r_0$ is the comoving coordinate and $a$ is the scale factor. $d\Omega_2^2$ expresses the 2-dimensional sphere with unit radius. $k=1$, $0$ and $-1$ corresponds to a closed, flat and open cosmology, respectively. Define $r=ar_0$, the metric ({\ref{eq4.1}) becomes

\begin{equation}
ds^2 = - \frac{1-r^2/r_A^2}{1-kr^2/a^2}dt^2 -\frac{2Hr}{1-kr^2/a^2}dtdr+\frac{1}{1-kr^2/a^2}dr^2+r^2d\Omega_2^2. \label{eq4.2}
\end{equation}

\noindent In the above equation, $r_A=\frac{1}{\sqrt{H^2+k/a^2}}$ is the location of the apparent horizon and $H=\frac{\dot a}{a}$ is the Hubble parameter. When $k=0$, the apparent horizon is just the Hubble horizon. In the section 2, the metric ({\ref{eq2.1}) describes the pure de Sitter spacetime when $M=0$. If we perform the transformation $d\tilde t=dt+\frac{Hr}{1-kr^2/a^2}dr$ on the metric (\ref{eq4.2}) and let $k=0$ and $r_A = H^{-1}=\sqrt{\frac{3}{\Lambda}}$, the metric ({\ref{eq4.2}) is reduced to the pure de Sitter metric. Therefore, the pure de Sitter spacetime is a special case of the FRW cosmology and there may exist some similar properties for them.

For convenience of calculation, we order $F(r)=\frac{1-r^2/r_A^2}{1-kr^2/a^2}$, $Y(r)=\frac{2Hr}{1-kr^2/a^2}$ and $G(r)=1-kr^2/a^2$. The temperature and the entropy at the apparent horizon are

\begin{equation}
T=\frac{1}{2\pi r_A}, \quad S=\frac{A}{4}=\pi r_A^2, \label{eq4.3}
\end{equation}

\noindent respectively \cite{CCH1}. The Misner-Sharp mass in the apparent horizon is gotten as $M=\frac{1}{2}r_A$ \cite{MS,CCH2,LZ,ZRS}. To discuss the thermodynamics, we first calculate the radial Einstein's equation at the apparent horizon. It is gotten as

\begin{equation}
G_r^r\mid_{r_A} = 8\pi T_r^r\mid_{r_A}= \frac{4 F^{\prime}(r_A)-Y^2(r_A)}{r_A^2Y^2(r_A)}. \label{eq4.4}
\end{equation}

\noindent Although the temperature has been expressed in Eq. ({\ref{eq4.3}), it isn't convenient for us to discuss the thermodynamics. We write the temperature in the following form

\begin{equation}
T=-\frac{\sqrt{G(r_A)}F^{\prime}(r_A)}{2\pi Y(r_A)}. \label{eq4.5}
\end{equation}

\noindent From Eqs. (\ref{eq4.4}) and (\ref{eq4.5}), the temperature is solved as

\begin{equation}
T=-\frac{\sqrt{G(r_A)}Y(r_A)}{8\pi r_A}\left(8\pi r_A^2 T_r^r\mid_{r_A}+1\right). \label{eq4.6}
\end{equation}

\noindent Multiplying by $\delta S$ on the both sides of the above equation yields

\begin{equation}
T\delta S= -2 r_A T_r^r\mid_{r_A} \delta S -\frac{1}{2}\delta r_A. \label{eq4.7}
\end{equation}

\noindent Move the first term on rhs to lrs. Identifying the energy as the negative Misner-Sharp mass yields $\delta E = -\frac{1}{2}\delta r_A$. Using the relation between the entropy and the horizon area, we get the modified first law of thermodynamics at the apparent horizon as

\begin{equation}
\delta E=T\delta S + \sigma \delta A, \label{eq4.8}
\end{equation}

\noindent where $\sigma = \frac{1}{2}r_A T_r^r\mid_{r_A}=-\frac{3}{16\pi r_A}$ denotes the surface tension at the apparent horizon. This formula is full in consistence with that derived in Eq. (\ref{eq3.6}) and shows the similarity between the de Sitter spacetime and the FRW cosmology. The corresponding Small relation

\begin{equation}
E= 2(TS + \sigma A), \label{eq4.9}
\end{equation}

\noindent is obeyed. If Eq. (\ref{eq4.8}) is written as $\delta E=T_{e}\delta S$, $T_{e} = -\frac{1}{4\pi r_A}$ denotes the effective temperature at the apparent horizon. The negative temperature appears and isn't allowed here. Therefore, when the heat flux produces at the horizon, the system is doing work on the the surroundings. The worked term exists in the first law at the apparent horizon.

We make further to explore the thermodynamics of the FRW cosmology. Replacing the area $A$ with the volume $V$ in Eq. (\ref{eq4.8}), we get

\begin{equation}
\delta E=T\delta S + P \delta V, \label{eq4.10}
\end{equation}

\noindent where $P=\frac{2\sigma}{r_A}=-\frac{3}{8\pi r_A^2}$ is the pressure at the apparent horizon. Now the corresponding Smarr relation isn't satisfied.

\section{Discussion and Conclusion}

When $M=0$, the Schwarzschild de Sitter black hole is reduced to the pure de Sitter spacetime which is the special case of the FRW spacetime. Therefore, the modified first laws of the de Sitter peacetime also satisfy Eqs. (\ref{eq3.6}) and (\ref{eq3.10}). In the FRW spacetime, if the apparent horizon radius $r_A$ is treated as a variable associated to the pressure $P=-\frac{3}{8\pi r_A^2}$, the modified law (\ref{eq4.10}) can be further written as $\delta E_0=T\delta S -  V\delta P$, where $ E_0=0$. The corresponding Smarr relation is satisfied. Furthermore, the laws (\ref{eq2.7}), (\ref{eq3.6}) and (\ref{eq4.8}) can be written as a relation $\delta E=\delta Q + \delta W$, where $\delta E$ express the change of the energy of the system, $\delta Q = T\delta S$ denotes the change of heat and $\delta W$ is a worked term. At the black hole horizon, the worked term in the modified law can disappear. While at the cosmological and apparent horizons, the worked terms must exist.

In this paper, we derived the modified first laws of thermodynamics at the black hole horizon and the cosmological horizon of the Schwarzschild de Sitter black hole and the apparent horizon of the FRW cosmology by the surface tensions, respectively. In the black hole, the black hole horizon and the cosmological horizon were seen as two independent thermodynamical systems. The modified laws (\ref{eq2.7}) and (\ref{eq3.6}) were gotten when the cosmological constant was fixed. When the constant was seen as a variable associated to the pressure, we obtained the first laws (\ref{eq2.11}) and (\ref{eq3.10}) in the extended phase spaces. The law at the the apparent horizon of the FRW spacetime takes on the same form as that at the cosmological horizon, but is different from that at the black hole horizon.

\vspace*{2.0ex}
\noindent \textbf{Acknowledgments}
This work is supported by the National Natural Science Foundation of China (Grant No. 11205125), by Sichuan Province Science Foundation for Youths (Grant No. 2014JQ0040) and by the Innovative Research Team in China West Normal University (Grant No. 438061).

\end{document}